\documentclass[aps,pra,twocolumn,groupaddress,showpacs]{revtex4}

\usepackage{dcolumn}
\usepackage{graphicx}
\usepackage{epsf}
\usepackage{amsmath}
\usepackage{amssymb}
\usepackage{bm}
\usepackage{times}

\newcolumntype{.}{D{x}{}{-1}}
\newcommand{\empt}{\multicolumn{1}{c}{\mbox{---}}}

\def\Za{{Z\,\alpha}}
\def\bfr{{\vec{r}}}

\def\bfsigma{{\vec{\sigma}}}

\newcommand{\onehalf}{{\textstyle{1\over 2}}}

\begin{document}
\preprint{Version 1.0}

\title{Quantum electrodynamic corrections\\
to the hyperfine structure of excited $S$ states}

\author{Ulrich D. Jentschura}
\affiliation{Max--Planck--Institut f\"ur Kernphysik,
Saupfercheckweg 1, 69117 Heidelberg, Germany}

\author{Vladimir A. Yerokhin}
\affiliation{Max--Planck--Institut f\"ur Kernphysik,
Saupfercheckweg 1, 69117 Heidelberg, Germany}
\affiliation{Center
for Advanced Studies, St.~Petersburg State Polytechnical
University, Polytekhnicheskaya 29, St.~Petersburg 195251, Russia}

\begin{abstract}
State-dependent quantum electrodynamic corrections are evaluated
for the hyperfine splitting of $nS$ states for arbitrary principal
quantum number $n$. The calculations comprise both the self-energy
and the vacuum-polarization correction of order
$\alpha\,(\Za)^2\,E_F$ and the recoil correction of order
$(\Za)^2\,(m/M)\,E_F$.  Higher-order corrections are summarized and
partly reevaluated as well. Accurate predictions for hydrogen
hyperfine splitting intervals of $nS$ states with $n = 2, \ldots,
8$ are presented. The results obtained are important due to steady
progress in hydrogen spectroscopy for transitions involving highly
excited $S$ states.
\end{abstract}


\maketitle

%
%
\section{Introduction}
\label{sec1}

Investigations of the hyperfine structure in light hydrogen-like
atomic systems are interesting for two main reasons. First,
accurate measurements of the hyperfine splitting (hfs), combined
with high-precision {\em ab initio} calculations, can yield
fundamental tests of bound-state QED theory. Second, the
accurate knowledge of the hfs also constitutes a necessary
ingredient in the determination of fundamental constants from
hydrogen and deuterium spectroscopy. The hyperfine components of
transitions in hydrogen can be accurately resolved at the current
level of spectroscopic accuracy, and the knowledge of the hfs of
excited states is therefore necessary for the interpretation of
the experimental data.

The ground-state hfs in hydrogen is known with an outstanding
accuracy (a part in $10^{12}$) for over of 3 decades, and the
value of
\begin{equation}
\label{E1S}
\Delta E_{1S} = 1\,420\,405\,751.768(1)\,{\rm Hz}
\end{equation}
has been obtained in Ref.~\cite{Ka2005} as a conservative average
of various experimental investigations of comparable accuracy, the
first of which was reported in
Ref.~\cite{EsDoBaHo1971}. Unfortunately, our
theoretical understanding of the ground-state hfs is limited by
the insufficient knowledge of the nuclear charge and magnetization
distributions, whose contribution of about $-50\,{\rm kHz}$ (30
ppm) cannot be accurately calculated at present.

One of the possibilities to overcome this difficulty \cite{St1963}
is to study the normalized difference of the $nS$ and $1S$ hfs
intervals,
\begin{equation}\label{Delta}
    \Delta_{n} = n^3 \, \Delta E_{nS} - \Delta E_{1S}\,.
\end{equation}
In this combination of energy intervals, the hfs energy shifts due
to the nuclear charge and magnetization distributions are largely
eliminated. Indeed, the lowest-order nuclear corrections to
$\Delta E_{1S}$ and $\Delta E_{nS}$ scale with the nonrelativistic
electron density at the position of the nucleus $|\phi_n(r=0)|^2$
which is strictly proportional to $n^{-3}$.
The nuclear effects thus do not contribute to the difference $\Delta_{n}$ to
leading order. Theoretical investigations show that the specific
difference $\Delta_{n}$ provides an opportunity to test the QED
theory of bound states on a level of about two orders of magnitude
better than for the ground-state hyperfine interval $\Delta
E_{1S}$ alone~\cite{Ka2005}. According to widely accepted
terminology, the corrections that depend on $n$ through
$|\phi_n(r=0)|^2$ only are called ``state independent''. Thus,
only state dependent correction should be considered in
theoretical investigations of the difference $\Delta_n$.

Accurate experimental results for the difference $\Delta_{2}$ are
presently available for the hydrogen, deuterium, and the $^3$He ion.
Notably, recent progress has been achieved for hydrogen
\cite{KoFiKaHa2004} and deuterium~\cite{KoFeKaHa2004} 
via optical spectroscopy, by comparing
the $1S$ and $2S$ hyperfine splittings via a
phase-coherent optical measurements of the $1S(F=0)
\Leftrightarrow 2S\,(F\!=\!0)$ vs.~the $1S(F\!=\!1) \Leftrightarrow
2S\,(F\!=\!1)$ transition. The best absolute accuracy for the
difference $\Delta_{2}$ is, however, still obtained for the
$^3$He ion in a combination of two relatively old
measurements~\cite{ScFoDe1969,PrWa1977},
\begin{equation}\label{1}
 \Delta_{2}(^3{\rm He}^+) = 1\,189.979\,(71)\,\,{\rm kHz}\,.
\end{equation}

While the specific difference of the $2S$ and $1S$ hfs intervals
has been a subject of experimental and theoretical
\cite{Zw1961,St1963} investigations for a long time, the
difference $\Delta_n$ for $n>2$ has attracted much less attention up
to now. The case $n>2$ is, however, becoming of significant interest nowadays,
due to steady progress in hydrogen spectroscopy for transitions
involving highly excited $S$ states. Two ongoing experiments could
be mentioned in this connection, which concern the hydrogen
$1S-3S$ transition~\cite{UdPriv2005,ArPriv2004} and are expected
to reach a sub-kHz level of accuracy.

In the present work, we perform a calculation of the leading
state-dependent self-energy and vacuum-polarization corrections
for an arbitrary $nS$ state. For the case $n=2$, we reproduce the
well-known results by Zwanziger \cite{Zw1961}. We also generalize
the derivation of the leading state-dependent recoil correction
given by Sternheim \cite{St1963} for $n=2$ to general $n$.
Next, we summarize and partly reevaluate the state-dependent
higher-order correction and present numerical results for the
difference $\Delta_n$ with $n=2,\dots, 8$ for hydrogen.

This paper is organized as follows: Basic quantities are introduced
in Sec.~\ref{sec2}. Third-order state-dependent corrections are
analyzed and summarized in Sec.~\ref{sec3}. Among these,
self-energy corrections are treated in Sec.~\ref{sec31},
vacuum-polarization corrections in Sec.~\ref{sec32}, and recoil
corrections in Sec.~\ref{sec33}. The current status of
higher-order state-dependent corrections is discussed in
Sec.~\ref{subsec42}. Finally, the total theoretical predictions
for the normalized difference of the hfs intervals $\Delta_n$ in
hydrogen are presented in Sec.~\ref{subsec43} for $n=2,\dots,8$.

%
%
\section{General formulas and notations}
\label{sec2}

We are using natural units with $\hbar = c = \epsilon_0 = 1$. The
electron charge is denoted by $e = -|e|$ and $\alpha =
e^2/(4\pi)$. The magnetic dipole moment of the nucleus is
\begin{equation} \label{magmoment}
 \vec{\mu} = g \,\mu_N\, \vec{I}\,,
\end{equation}
where $g$ denotes the nuclear $g$ factor, $\mu_N = |e|/(2 m_p)$ is
the nuclear magneton, and $m_p$ is the proton mass. The vector
potential generated by the nuclear dipole moment is
\begin{equation}\label{Ahfs}
 \vec{A} = \frac{\vec{\mu} \times \vec{r}}{4 \, \pi \,
r^3} = -\frac{\vec{\mu}}{4\pi} \times \vec{\nabla} \frac{1}{r} \,.
\end{equation}
The interaction of the bound electron with the dipole nuclear
magnetic field is given by the Fermi-Breit operator,
\begin{equation}
\label{Vhfs} V_{\rm hfs} = -e \, \vec{\alpha}\cdot \vec{A} =
\frac{|e|}{4\pi}\, \frac{\vec{\alpha} \cdot (\vec{\mu} \times
\vec{r})}{r^3}\,.
\end{equation}
The expectation value of the Fermi-Breit operator on Dirac
point-nucleus wave functions is well-known. We write it as
\begin{align}\label{Breit1}
    E_{\rm hfs} = &\ \alpha\, (\Za)^3\,\frac{g}{2}
    \frac{m^2}{m_p}
    \frac{\kappa}{|\kappa|}
    \frac1{n^3\,(2\kappa+1)(\kappa^2-1/4)}\,
       \nonumber \\ &\times
    A(\Za)\,[F(F+1)-I(I+1)-j(j+1)]\,,
\end{align}
where  $A(\Za)$ is a relativistic factor $[A(\Za) = 1 + {\cal
O}(\Za)^2]$,
\begin{equation}\label{Breit2}
    A(\Za) = n^3\, |\kappa| (2\kappa+1)\,
      \frac{2\kappa(\gamma+n_r)-N}{N^4\,\gamma\,(4\gamma^2-1)}\,.
\end{equation}
Here, $N = \sqrt{n_r^2+2n_r\gamma+\kappa^2}$, $n_r = n -
|\kappa|$, $\gamma = \sqrt{\kappa^2 - (Z\alpha)^2}$, $n$ is the principal
quantum number of the electron, $\kappa$ is its Dirac angular
quantum number, $j = |\kappa|-1/2$ is the total momentum of the
electron, and $m$ is the electron mass.

For future reference, we also give the magnetic field
corresponding to the vector potential (\ref{Ahfs}),
\begin{equation}
\vec{B} = \vec{\nabla} \times \vec{A} =
\frac23\, \vec{\mu}\, \delta^3(r) +
\frac{3 (\vec{\mu} \, \cdot \, \hat{\vec{r}}) \, \hat{\vec{r}} -
\vec{\mu} }{4 \pi \, r^3}\,.
\end{equation}

In the nonrelativistic limit, the hyperfine Hamiltonian $H_{\rm
hfs}$ is given by the sum of two terms, the first of which is
proportional to $\vec{\sigma}\cdot \vec{B}$ and is denoted here as
$H_S + H_D$, whereas the second one (labeled $H_L$) corresponds to
the interaction of the nuclear moment with the magnetic field of
the moving electron, which in turn is proportional to the orbital
angular momentum $\vec{L}$. We have
\begin{subequations}
\label{Hhfs}
\begin{align}
H_{\rm hfs} =& \; H_S + H_D + H_L \,,
\\
\label{HS} H_S =& \; \frac{|e|}{3 m} \, \vec{\sigma}\cdot
\vec{\mu} \,\, \delta^3(r) \,,
\\
\label{HD} H_D =& \; \frac{|e|}{8 m} \,  \frac{ 3 \,\,
\vec{\sigma} \cdot \hat{\vec{r}} \, \, \vec{\mu} \cdot
\hat{\vec{r}} - \vec{\sigma} \cdot \vec{\mu}}{\pi r^3} \,,
\\
\label{HL}
H_L =& \; \frac{|e|}{4 m} \,
\frac{\vec{\mu}\cdot \vec{L}}{\pi \, r^3}\,.
\end{align}
\end{subequations}
Here, $\hat{\vec{r}}$ is the unit vector in the direction of
$\vec{r}$. For the Schr\"{o}dinger wave function  $\phi_n$ of an
$nS$ state, the expectation value of the nonrelativistic
Hamiltonian is
\begin{align}
\left< H_{\rm hfs} \right> =  \left< H_S \right>=
\frac{|e|}{3m}\,\left<\bfsigma\cdot\vec{\mu}\right>\,|\phi_n(0)|^2\,,
\end{align}
and the splitting between the ground-state levels with $F = I +
\onehalf$ and $F = I - \onehalf$ gives us the Fermi energy
\begin{equation}\label{Fermi}
E_F = \frac{|e|}{3m}\,g \,\mu_N\,|\phi_{n=1}(0)|^2 \,(2 I+1)\,,
\end{equation}
where $|\phi_{n=1}(0)|^2 = (\Za)^3 m^3/\pi$ in the
non-recoil limit.

%
%
\section{THIRD--ORDER CORRECTIONS}
\label{sec3}

%
%
\subsection{Self--energy}
\label{sec31}

The leading state-dependent self-energy correction to the
hyperfine splitting can be conveniently expressed as
\begin{equation}
\label{defSE}
\delta \Delta^{\rm SE}_n = \frac{\alpha}{\pi}(\Za)^2 E_F
\left\{ a^{\rm SE}_{21}(n,1) \, \ln[(Z\alpha)^{-2}]
+ a^{\rm SE}_{20}(n,1) \right\}.
\end{equation}
Here, $\delta \Delta^{\rm SE}_n$ is the contribution to the
normalized difference $\Delta_n$ due to self-energy effects, where
$\Delta_n$ is defined according to Eq.~(\ref{Delta}). In general,
we will denote various contributions to $\Delta_n$ by 
the symbol $\delta \Delta_n$ with appropriate superscripts. 
The coefficients $a^{\rm SE}_{ij}(n,1)$ are understood
as originating from the difference $a^{\rm SE}_{ij}(n,1) = a^{\rm
SE}_{ij}(nS) - a^{\rm SE}_{ij}(1S)$, with $a^{\rm SE}_{ij}(nS)$
being the corresponding coefficient for the $nS$ state. As usual,
the first index of $a^{\rm SE}_{ij}$ counts the power of
$Z\alpha$, and the second one indicates the power of the logarithm
$\ln[(Z\alpha)^{-2}]$.

The self-energy correction~(\ref{defSE}) consists of two parts
induced by the low-energy and the high-energy virtual
photons~\cite{Pa1995}. The low-energy part can be immediately
obtained by generalizing formulas given in
Refs.~\cite{Pa1995,NiKi1997,Je2003jpa}. The corresponding
contribution expressed in units of $\alpha (\Za)^2/\pi$ reads:
\begin{align}
& \frac{{\cal C}_L}{\frac{\alpha}{\pi}\,(\Za)^2} = \frac83 \,
\left[ \frac34 - \frac1n + \frac{1}{4 n^2} + \gamma + \Psi(n) -
\ln(n) \right]
\nonumber\\
& \quad \times \ln\left( \frac{\epsilon}{(\Za)^2 \, m} \right) +
N(nS) - N(1S)\,.
\end{align}
Here, $N(nS)$ is a delta correction to the Bethe logarithm, whose
numerical values are given in Table~\ref{NnS}.

\begin{table}[htb]
\begin{center}
\begin{minipage}{8.6cm}
\begin{center}
\caption{\label{NnS} Numerical values of the quantity $N(nS)$.}
\begin{tabular}{cc}
\hline \hline
n & $N(nS)$ \\
\hline
1 & 17.855\,672\,03(1) \\
2 & 12.032\,141\,58(1) \\
3 & 10.449\,809(1) \\
4 &  9.722\,413(1) \\
5 &  9.304\,114(1) \\
6 &  9.031\,832(1) \\
7 &  8.840\,123(1) \\
8 &  8.697\,639(1) \\
\hline \hline
\end{tabular}
\end{center}
\end{minipage}
\end{center}
\end{table}

We now turn to the contribution due to high-energy virtual
photons. Up to relative order $\alpha (\Za)^2$, we can
use the modified Dirac Hamiltonian $H_{\rm rad}$ (for a derivation
see, e.g., Chap.~7 of~\cite{ItZu1980}), which reads
\begin{align}
\label{HDm}
H_{\rm rad} =& \; \vec{\alpha} \cdot
\left[\vec{p} -{\mathrm e} \, F_1(\vec{\nabla}^2) \, \vec{A}\right]
+ \beta\,m + F_1(\vec{\nabla}^2) \, V
\nonumber\\
& + F_2(\vec{\nabla}^2) \, \frac{e}{2\,m} \, \left({\mathrm i}\,
\vec{\gamma} \cdot \vec{E} - \beta \, \vec{\sigma} \cdot \vec{B}
\right)\,.
\end{align}
This Hamiltonian leads to various self-energy corrections. The
first of these is an $F_2(0)$ correction to the effective
potential, evaluated on the relativistic wave functions. It is
generated by the following term in Eq.~(\ref{HDm}),
\begin{equation}
\label{deltaHhfs} \delta H = -F_2(0) \, \frac{e}{2 m} \, \beta \,
\vec{\sigma} \cdot \vec{B} = \frac{\alpha}{2 \pi} \, \beta \, (H_S
+ H_D)\,,
\end{equation}
where the Schwinger result $F_2(0) = \alpha/(2\pi)$ has been used,
$H_S$ and $H_D$ are given in Eqs.~(\ref{HS}) and~(\ref{HD}),
respectively, and $\beta$ is the Dirac $\gamma^0$ matrix in the
Dirac representation. The corresponding relative correction to the
Fermi energy (\ref{Fermi}) is
\begin{equation}
\label{C1pre}
\frac{\alpha}{2\pi} \,
\frac{\left< \psi \left| \beta \, (H_S + H_D) \right| \psi \right>}
{\left< \phi \left| H_S \right| \phi \right>}\,.
\end{equation}
Here, $\psi$ is the fully relativistic (Dirac) hydrogen wave
function expanded in powers of $\Za$, whereas $\phi$ is the
nonrelativistic (Schr\"{o}dinger--Pauli) counterpart. Under the
replacement $\psi \to \phi$, Eq.~(\ref{C1pre}) simply gives the
leading term $\alpha/(2\pi)$. The numerator of Eq.~(\ref{C1pre})
diverges in relative order $(\Za)^2$ when evaluated on an $nS$
state. A finite result is obtained, however, when the weighted (or
normalized) difference of matrix elements is considered. We define
the normalized difference for the general operator $A$ as
\begin{equation}
\left< \! \left< A \right> \! \right> =
n^3 \left< nS | A | nS \right> - \left< 1S | A | 1S \right> \,.
\end{equation}
The correction (\ref{C1pre}) leads to the following contribution
to the normalized difference (\ref{Delta}) of hfs intervals,
\begin{equation}
{\cal C}_1 =
\frac{\delta \Delta_n^{{\rm SE},1}}{E_F} =
\frac{\alpha}{2\pi} \,
\frac{\left< \! \left< \psi \left|
\beta \, (H_S + H_D) \right| \psi \right> \! \right>}
{\left< \phi \left| H_S \right| \phi \right>}\,.
\end{equation}

The second correction (${\cal C}_2$) is an
$F_2'$ correction to the effective potential
(\ref{deltaHhfs}), i.e.
\begin{equation}
-F'_2(0) \, \frac{e}{2m} \,
\beta \, \vec{\nabla}^2 \, \vec{\sigma}\cdot\vec{B}\,,
\end{equation}
to be evaluated on the nonrelativistic wave functions. For the
third correction ${\cal C}_3$, we have to evaluate an $F_1'$
correction to the effective potential (\ref{HS}); the relevant
Hamiltonian can be expressed as $F'_1(0) \, \vec{\nabla}^2 H_S$.
The forth correction is a second-order correction due to an
effective one-loop Lamb-shift potential, which can be expressed as
\begin{align}
\Delta V =& \; \alpha\, (Z\alpha) \, \left[ \frac43\, \ln
\left(\frac{m}{2\,\epsilon}\right) + \frac{10}{9} \right] \,
\frac{\delta^3(r)}{m^2}
\nonumber\\
=& \; \frac{\alpha}{3 \pi}\, (Z\alpha) \, \left[ \ln
\left(\frac{m}{2\,\epsilon}\right) + \frac{5}{6} \right] \,
\frac{\vec{\nabla}^2}{m^2} V\,.
\end{align}
Here, $\epsilon$ is a noncovariant low-energy photon cut-off and
$V$ denotes the Coulomb potential $V = - \Za/r$. Finally, the
fifth correction is a second-order contribution due to
negative-energy states and is induced by the relativistic
hyperfine potential $V_{\rm hfs}$ as given in Eq.~(\ref{Vhfs}) and
the term
\begin{equation}
F_2(0) \, \frac{e}{2\,m} \, {\mathrm i}\,
\vec{\gamma} \cdot \vec{E}
\end{equation}
from the modified Dirac Hamiltonian (\ref{HDm}), where $\vec{E}$
is the electric field generated by the Coulomb potential.
From the $r$-scaling of the two involved Hamiltonians, it is clear that 
the resulting operator has to be proportional to $1/r^4$. 
The prefactor can be obtained using Dirac algebra and 
considering the fact that the main contribution comes from 
negative-energy states with an energy $\approx -m$.

The high-energy corrections discussed so far are explicitly given
by
\begin{subequations}
\begin{align}
\label{C1}
{\cal C}_1 =& \frac{\alpha}{2\,\pi}\,
\frac{\left< \! \left< \psi \left|
\beta \, (H_S + H_D) \right| \psi \right> \! \right>}
{\left< \phi \left| H_S \right| \phi \right>}\,,
\\
\label{C2}
{\cal C}_2 =& \frac{\alpha}{12\,\pi}\,
\frac{ \left< \! \! \left< \frac{\vec{\nabla}^4}{m^4} V
\right> \! \! \right> }
{ \left< \frac{\vec{\nabla}^2}{m^2} V \right> } \,,
\\
\label{C3}
{\cal C}_3 =& \frac{\alpha}{3\,\pi}\,
\left[ \ln\left( \frac{m}{2\,\epsilon} \right) + \frac{11}{24} \right]\,
\frac{ \left< \! \! \left< \frac{\vec{\nabla}^4}{m^4} V
\right> \! \! \right> }
{ \left< \frac{\vec{\nabla}^2}{m^2} V \right> } \,,
\\
\label{C4}
{\cal C}_4 =& \frac{2 \alpha}{3\,\pi}\,
\left[ \ln\left( \frac{m}{2\,\epsilon} \right) + \frac{5}{6} \right]\,
\frac{ \left<\!\!\left< \frac{\vec{\nabla}^2}{m^2} V
\, \frac{1}{(E-H)'} \, \frac{\vec{\nabla}^2}{m^2} V
\right> \!\! \right>}
{ \left< \frac{\vec{\nabla}^2}{m^2} V \right> }\,,
\\
\label{C5}
{\cal C}_5 =& \frac{\alpha}{\pi}\,
\frac{ \left< \!\! \left< \frac{\alpha^2}{2 \, m^3 \, r^4}
\right> \!\! \right>}
{ \left< \frac{\vec{\nabla}^2}{m^2} V \right> }\,.
\end{align}
\end{subequations}
Here, we reemphasize that $| \psi \rangle$ is the relativistic
wave function, $| \phi \rangle$ is the nonrelativistic wave
function, and all matrix elements $\left< A \right>$, by default,
are understood in terms of the nonrelativistic wave function.

\begin{table}[htb]
\begin{center}
\begin{minipage}{8.6cm}
\begin{center}
\caption{\label{a20} Numerical values of the nonlogarithmic
self-energy coefficient for the normalized difference [$a^{\rm
SE}_{20}(n,1)$] and for the single $nS$ states [$a^{\rm
SE}_{20}(nS)$] in the range $n = 1,\dots,8$.}
\begin{tabular}{c@{\hspace{0.5cm}}c@{\hspace{0.5cm}}c}
\hline
\hline
n & $a^{\rm SE}_{20}(n,1)$ & $a^{\rm SE}_{20}(nS)$ \\
\hline
1 &      \empt           &     17.122~338~75(1) \\
2 &     $-$5.221~233~33(1) &     11.901~105~41(1) \\
3 &       $-$6.705~291(1)  &       10.417~048(1) \\
4 &       $-$7.402~951(1)  &        9.719~388(1) \\
5 &       $-$7.809~635(1)  &        9.312~703(1) \\
6 &       $-$8.076~773(1)  &        9.045~565(1) \\
7 &       $-$8.266~081(1)  &        8.856~258(1) \\
8 &       $-$8.407~461(1)  &        8.714~878(1) \\
\hline
\hline
\end{tabular}
\end{center}
\end{minipage}
\end{center}
\end{table}

The results for the normalized $S$-state difference, expressed in
units of $\alpha (\Za)^2/\pi$, are:
\begin{subequations}
\begin{align}
\frac{{\cal C}_1}{\frac{\alpha}{\pi}\,(\Za)^2} =& \; \frac{19}{48}
+ \frac58 - \frac{49}{48\,n^2} - \frac14\, \left[ \gamma + \Psi(n)
- \ln(n) \right] \,,
\\
\frac{{\cal C}_2}{\frac{\alpha}{\pi}\,(\Za)^2} =& \;
\frac{1}{6} \, \left( \frac{1}{n^2} - 1\right)\,,
\\
\frac{{\cal C}_3}{\frac{\alpha}{\pi}\,(\Za)^2} =& \;
\frac{1}{6} \, \left( \frac{1}{n^2} - 1\right) \,
\left[ \frac23\, \ln\left( \frac{m}{2 \epsilon}\right) +
\frac{11}{36} \right]\,,
\\
\frac{{\cal C}_4}{\frac{\alpha}{\pi}\,(\Za)^2} =& \;
\frac83
\left[ \ln\left( \frac{m}{2 \epsilon}\right) + \frac{5}{6} \right]
\nonumber\\
& \; \times \left[ 1 - \frac{1}{n} + \gamma + \Psi(n) - \ln(n)
\right] \,,
\\
\frac{{\cal C}_5}{\frac{\alpha}{\pi}\,(\Za)^2} =& \;
-\frac23 + \frac{1}{2 n} + \frac{1}{6 n^2}
+ \gamma + \Psi(n) - \ln(n) \,.
\end{align}
\end{subequations}

Adding all the contributions together, we obtain the following
result for the self-energy correction (\ref{defSE}),
\begin{align}
& a^{\rm SE}_{21}(n,1)
\ln[(Z\alpha)^{-2}] + a^{\rm SE}_{20}(n,1)
= \frac{{\cal C}_L + \sum_{j=1}^5 {\cal C}_j}
{\frac{\alpha}{\pi}\,(\Za)^2}\,.
\end{align}
Of course, the dependence on the noncovariant photon energy cutoff
$\epsilon$ disappears in the final answer. The result for the 
logarithmic term is~\cite{Ka2001hyp}
\begin{equation}
a^{\rm SE}_{21}(n,1) = \frac83 \, \bigg[ \frac34 - \frac1n +
\frac{1}{4 n^2} + \gamma + \Psi(n) - \ln(n) \bigg] \,.
\end{equation}
For the nonlogarithmic term $a^{\rm SE}_{20}(n,1)$, we obtain the
general result
\begin{align}
& a^{\rm SE}_{20}(n,1) = \;
N(nS) - N(1S)
\nonumber\\
& +\frac{71}{48}
- \frac{79}{72\,n}
- \frac{55}{144\,n^2}
+ \frac{107}{36} \, \left[ \gamma + \Psi(n) - \ln(n) \right]
\nonumber\\
& \; - \frac83 \, \ln(2) \, \left[ \frac34 - \frac1n + \frac{1}{4
n^2} + \gamma + \Psi(n) - \ln(n) \right] \,. \label{seres}
\end{align}
In the particular case $n=2$, we reproduce the known value for
this coefficient~\cite{B5}. Explicit numerical results for $a^{\rm
SE}_{20}(n,1)$ are given in Table~\ref{a20} for $n = 1,\dots,8$.
In the table, we also list the values of $a^{\rm SE}_{20}(nS)$
obtained with the help of an improved $1S$ numerical value, which
we give here for reference purposes,
\begin{equation}
\label{a20root} a^{\rm SE}_{20}(1S) = 17.122\,338\,75(1)\,.
\end{equation}
This result can be immediately obtained according to the improved
numerical evaluation of the low-energy part as described in
Ref.~\cite{Je2003jpa}, which contains a correction to the Bethe
logarithm induced by a Dirac-delta local potential (see also the
entries in the forth column of Table~II of
Ref.~\cite{JeCzPa2005}).

%
%
\subsection{Vacuum polarization}
\label{sec32}

The leading state-dependent
vacuum-polarization correction to the hyperfine
splitting can be conveniently expressed as
\begin{equation}
\label{defVP} \delta \Delta^{\rm VP}_n =
\frac{\alpha}{\pi}\,(\Za)^2\,E_F\, a^{\rm VP}_{20}(n,1)\,.
\end{equation}
The correction
$\delta \Delta^{\rm VP}_n$ consists of two parts \cite{Zw1961}, with the
first one given by a matrix element of the radiatively corrected
external magnetic field and the other by a matrix element of the
vacuum-polarization operator between the wave functions corrected
by the presence of the external magnetic field.

We start with the first part. To the leading order, the
radiatively corrected magnetic interaction
(magnetic loop) is well-known to be
\begin{align}
\label{vp1}
& V_{\rm VP, mag}(\bfr) = V_{\rm hfs}(\bfr)
\nonumber\\
& \; \times \frac{2\alpha}{3\pi}\, \int_1^{\infty}dt\,
\frac{\sqrt{t^2-1}}{t^2}\, \left(1+\frac1{2t^2}\right)\, ( 1+2m
rt)\, e^{-2m rt}\,.
\end{align}
We recall that the matrix element of $V_{\rm hfs}$
between the Dirac wave functions is, for $nS$ states,
\begin{equation}  \label{vp2}
\left<n| V_{\rm hfs} |n \right> =
-\frac{E_F}{m^2 (\Za)^3}\, \int_0^{\infty}dr\,
  g_n(r)\, f_n(r)\,,
\end{equation}
where $g_n$ and $f_n$ are the upper and the lower radial component
of the Dirac wave function, respectively. We thus immediately have that
\begin{align}
\label{vp3} & \delta E^{\rm VP, mag}_n = \left<n| V_{\rm VP, mag}
|n \right>
\nonumber\\
& = -\frac{E_F}{m^2 (\Za)^3}\, \frac{2\alpha}{3\pi}\,
\int_1^{\infty}dt\, \frac{\sqrt{t^2-1}}{t^2}\,
\left(1+\frac1{2t^2}\right)\,
\nonumber\\
& \quad \times \int_0^{\infty}dr\, ( 1+2m rt)\, e^{-2m rt}\,
g_n(r)\, f_n(r)\,.
\nonumber \\
\end{align}
To the leading order in $\Za$ for an $nS$ state,
\begin{equation}  \label{vp5}
g_n(r) = \frac2n\, \left( \frac{\beta}{n}\right)^{3/2} e^{-\beta
r/n}\, {\cal L}^1_{n-1}\left( \frac{2\beta r}{n}\right)\,,
\end{equation}
and
\begin{equation}  \label{vp4}
f_n(r) = \frac1{2m}\, \frac{d}{dr}\, g_n(r)\,,
\end{equation}
where $\beta = \Za \,m$, and ${\cal L}^1_{n-1}$ are generalized
Laguerre polynomials. Performing the integration over $r$ in
Eq.~(\ref{vp3}) with help of entry (2.19.14.6)
in Vol.~2 of Ref.~\cite{PrBrMa2002}, expanding the result in
$\Za$, and integrating over $t$, we obtain
\begin{equation}\label{vp6}
\delta E^{\rm VP, mag}_n = \frac{E_F}{n^3}\, \frac{\alpha}{\pi}\,
(\Za)\, \left[ \frac{3\pi}{8} -
\frac2{15}\left(5+\frac1{n^2}\right)\,(\Za) \right]\,.
\end{equation}
The corresponding contribution to $\Delta_n$ is
\begin{align}
\label{vp7} \delta \Delta_n^{\rm VP, mag} = \frac{\alpha}{\pi}\,
(\Za)^2\, E_F\, \frac2{15}\left(1-\frac1{n^2}\right)\,.
\end{align}

The second vacuum-polarization contribution is given by the
second-order correction,
\begin{equation}\label{vp8}
  \delta E^{\rm VP, el}_n = 2 \, \left< n\left| V_{\rm hfs}\,
  \frac1{(E-H)^{\prime}}\, V_{\rm VP}\, \right| n\right>\,,
\end{equation}
where $V_{\rm VP}$ is the vacuum-polarization potential. Due to
spherical symmetry of $V_{\rm VP}$, only the $nS$ intermediate
states contribute in the above expression. To the leading order,
we have
\begin{equation}\label{vp9}
    V_{\rm VP}(\bfr) = -\frac4{15} \frac{\alpha\, (\Za)}{m^2}\,
    \delta(\bfr)\,,
\end{equation}
and we can replace $V_{\rm hfs} \to H_S$, with $H_S$ being given
in Eq.~(\ref{HS}). The second-order matrix element (\ref{vp8})
diverges for $nS$ states. It is, however, finite for the
normalized difference, with the result
\begin{equation}\label{vp11}
    \delta \Delta_n^{\rm VP, el} = -\frac8{15}\, E_F\,
      \frac{\pi\alpha}{m^5 (\Za)^2}\,
      \left<\!\!\left< \delta(\bfr)\, \frac1{(E-H)^{\prime}}\,
      \delta(\bfr) \right>\!\!\right>\,.
\end{equation}
Using the formulas from Ref.~\cite{JeCzPa2005} for the matrix
element, we arrive at
\begin{align}\label{vp12}
\delta \Delta_n^{\rm VP, el} =& \frac{\alpha}{\pi}\, (\Za)^2\,E_F
\,
\nonumber\\
& \times \left(-\frac8{15}\right)
\left[ 1 -\frac1n +\gamma + \Psi(n)-\ln (n) \right]\,.
\end{align}

Finally, the total result for the vacuum-polarization correction
[Eq.~(\ref{defVP})] reads
\begin{align} \label{vpres}
a_{20}^{\rm VP}(n,1) = -\frac8{15}
\left[ \frac34 -\frac1n+ \frac1{4 n^2} +\gamma +
\Psi(n)-\ln(n) \right]\,,
\end{align}
in agreement with Ref.~\cite{Ka2001hyp}.

%
%
%
\subsection{Recoil corrections}
\label{sec33}

The leading-order state-dependent recoil correction can be
parameterized as
\begin{equation}
\label{defREC}
\delta \Delta^{\rm REC}_n = (\Za)^2\,\frac{m}{M} \, E_F\,
a^{\rm REC}_{20}(n,1) \,,
\end{equation}
where $M$ is the mass of the nucleus. The general expression for
this correction was derived by Sternheim~\cite{St1963}.  It reads
\begin{equation}\label{rec1}
\delta E_n^{\rm REC} =
    \left< H_M^{(3)}\right>
    + \left< \left[2 H_M^{(1)}+H_M^{(2)}\right]
             \frac1{(E-H)^{\prime}} H_M^{(2)}\right>\,,
\end{equation}
where
\begin{subequations}
\begin{align}
\label{rec2}
H_M^{(1)} =& -\frac{e}{8m^2}\,\vec{\nabla}\cdot\vec{E}
-\frac{p^4}{8m^3} - \frac{e}{m}\,\vec{p}\cdot\vec{\delta A}\,,
\\
\label{rec3}
H_M^{(2)} =& -\frac{e}{2m}\,\vec{\sigma}\cdot\vec{B}\,,
\\
\label{rec4}
H_M^{(3)} =& -\frac{e}{2m}\,
\vec{\sigma}\cdot
\Bigl\{ -\frac{p^2}{4m^2}\,\vec{B} -\vec{B}\,\frac{p^2}{4m^2}
\nonumber \\
& - \frac{e}{2m}\, \vec{E}\times \vec{A}
+ \frac1{4m}\,(\vec{\delta E}\times \vec{p}- \vec{p}\times
\vec{\delta E})
\nonumber\\
& -\frac{\rm i}{8mM} \left[(\vec{p}\times
\vec{A}-\vec{A}\times\vec{p}),p^2\right] \Bigr\}\,.
\end{align}
\end{subequations}
Here, $\vec{A}$ is given in Eq.~(\ref{Ahfs}), $\vec{\delta E}$ is
the electric field induced by the scalar potential of a moving
magnetic dipole $\delta V$,
\begin{equation}\label{rec5}
\delta V =
-\frac{e}{4\pi}\,\left(\vec{\mu}+\frac{Ze}{2M}\,\vec{I}\right)\times
\frac{\vec{p}}{M} \cdot \vec{\nabla}\frac1r\,,
\end{equation}
and $\vec{\delta A}$ is the vector potential of the moving
nucleus,
\begin{equation}\label{rec6}
   e\,\vec{\delta A} = \frac1{8\pi}\, \frac{\Za}{Mr}\,
   \left(\vec{p}+\frac{\vec{r}}{r}\,\frac{\vec{r}}{r}\cdot\vec{p}\right)\,.
\end{equation}
The matrix elements in Eq.~(\ref{rec1}) diverge for $nS$ states, but
they yield a finite result for the normalized difference
$\Delta_n$, which reads
\begin{align} \label{recres}
& a_{20}^{\rm REC}(n,1) = -\frac32\,\left(1-\frac1{n^2}\right)
\nonumber\\
& -\frac{7\eta}{8}\left[ \frac{17}{28}-\frac{9}{14n}+\frac1{28n^2}
+\gamma +\Psi(n) -\ln(n) \right]
\nonumber\\
& +\left(\frac{1 - \eta}{2\eta} \right)
\left[ -\frac{11}{12} + \frac1{2n} + \frac{5}{12n^2}
+\gamma +\Psi(n) -\ln(n) \right]
\Biggr\}\,,
\end{align}
where $\eta = g M/(Z m_p)$ and $m_p$ is the proton mass. For
the particular case $n=2$, our result is in agreement with the one
originally obtained by Sternheim \cite{St1963}.

%
%
%
\subsection{Summary of the theory up to third order}
\label{subsec41}

To the leading order in the parameters $\alpha$, $\Za$, and $m/M$,
the normalized difference of the hyperfine-structure $nS$
intervals $\Delta_n$ is given by the sum of the relativistic
(Breit), self-energy, vacuum-polarization, and recoil corrections:
\begin{align} \label{the1}
\Delta_n = &\ (\Za)^2\,E_F\,\Bigl\{
    a_{20}^{\rm Br}(n,1)
    +\frac{\alpha}{\pi}\,\Bigl[ a_{21}^{\rm SE}(n,1)\, \ln[(\Za)^{-2}]
       \nonumber \\ & \
       +a_{20}^{\rm SE}(n,1)+a_{20}^{\rm VP}(n,1) \Bigr]
    + \frac{m}{M}\,a_{20}^{\rm REC}(n,1) \Bigr\}\,,
\end{align}
where the Fermi energy $E_F$ is defined as the splitting between
the ground-state levels with the atomic angular momentum $F =
I+1/2$ and $F = I-1/2$ calculated within the nonrelativistic
approximation and is given by
\begin{equation}\label{the2}
    E_F = \frac{4}{3}\,\alpha\,(\Za)^3\,\frac{m^2}{m_p}\,
    \frac{\mu}{\mu_N}\,
      \frac{2I+1}{2I} \left( 1+\frac{m}{M}\right)^{-3} \,,
\end{equation}
with the nuclear magnetic moment $\mu = g\,\mu_N\,I$. Notice that
this expression follows from Eq.~(\ref{Fermi}) after restoring the
correct reduced-mass dependence.

For the particular (and the most important) case $n=2$, the
coefficients in Eq.~(\ref{the1}) were obtained long ago
\cite{Br1930,Zw1961,St1963}. The full $n$ dependence of the
coefficients $a_{21}^{\rm SE}$ and $a_{20}^{\rm VP}$ was reported
in Ref.~\cite{Ka2001hyp}. In the present investigation, we 
have derived
the results for all coefficients in Eq.~(\ref{the1}) for 
general $n$. The self-energy, vacuum-polarization, and recoil
correction are given by Eqs.~(\ref{seres}), (\ref{vpres}), and
(\ref{recres}), respectively. The remaining second-order Breit
contribution to $\Delta_n$ is given by
\begin{equation}
a_{20}^{\rm Br}(n,1) = \: \left( \frac{1}{3} + \frac{3}{2\,n} -
\frac{11}{6\,n^2} \right)\,.
\end{equation}
%

%
%
\section{Higher--order corrections} \label{subsec42}

Higher-order QED and nuclear corrections to the difference
$\Delta_2$ were extensively investigated during the last years
\cite{Ka1997nuc,Ka2001hyp,YeSh2001hyp,KaIv2002,%
KaIv2002plb,YeArShPl2005,KaIv2005}. The general $n$ dependence of
the difference $\Delta_n$ received significantly less attention up
to now. In this section, we would like to summarize the results for
higher-order corrections and reevaluate some of them.

The higher-order relativistic (Breit) corrections are immediately
obtained by expanding the general formula (\ref{Breit2}):
\begin{align}
\frac{\delta \Delta_n^{\rm Br}}{E_F} = \:
&(Z\alpha)^4 \, \left(
\frac{25}{36} + \frac{25}{8\,n} - \frac{67}{36\,n^2}-
\frac{55}{12\,n^3}+ \frac{21}{8\,n^4} \right)
\nonumber \\
+ &
(Z\alpha)^6 \, \left(
 \frac{245}{216}+ \frac{245}{48\,n}- \frac{721}{432\,n^2}
 -  \frac{1195}{144\,n^3}
   \right. \nonumber \\
&\left.
 - \frac{33}{16\,n^4}+ \frac{147}{16\,n^5}
 - \frac{163}{48\,n^6}\right)
 \,,
\end{align}
where the sixth-order contribution is included for completeness.

The state-dependent two-loop correction to order
$\alpha^2\,(Z\,\alpha)^2$ was found in Ref.~\cite{Ka2001hyp} in
the logarithmic approximation. This result can be easily derived
if we observe that the leading one-loop $a_{10}$ correction for
the ground-state hfs is generated by an effective magnetic
form-factor correction [Eq.~(\ref{C1})] to the Hamiltonian
(\ref{HS}). We thus employ 
(\ref{HS}) as an input for a Dirac-delta correction to the Bethe
logarithm and obtain the result
\begin{align}
& \delta \Delta_n^{\mbox{\scriptsize two-loop}} =
\left(\frac{\alpha}{\pi}\right)^2 \, (Z\alpha)^2 \, E_F \,
\ln[(Z\alpha)^{-2}]\,
\nonumber\\
& \quad \times \frac43 \, \bigg[ \frac34 - \frac1n + \frac{1}{4
n^2} + \gamma + \Psi(n) - \ln(n) \bigg] \,,
\end{align}
in agreement with Ref.~\cite{Ka2001hyp}.

According to Ref.~\cite{Ka2001hyp}, analogous considerations are
valid also for the radiative-recoil correction, and hence
\begin{align}
&  \delta \Delta_n^{\mbox{\scriptsize rad-rec}}
= \frac{\alpha}{\pi} \,
(Z\alpha)^2 \, \frac{m}{M} \, E_F \, \ln[(Z\alpha)^{-2}]\,
\nonumber\\
& \quad \times
\left(-\frac{16}{3}\right) \, \bigg[ \frac34 - \frac1n +
\frac{1}{4 n^2} + \gamma + \Psi(n) - \ln(n) \bigg] \,.
\end{align}

We now turn our attention to the state-dependent recoil correction
to order $(m/M)\,(\Za)^3\,E_F$, which we evaluate in the
logarithmic approximation. We have identified two such
contributions. The first one can be obtained as a second-order
perturbation correction induced by two effective local potentials,
the first one being $H_S$ [Eq.(\ref{HS})] and the second one
corresponding to the logarithmic recoil correction to the Lamb
shift to order $(\Za)^5\,m^2/M$. The result is
\begin{align} \label{HRECa}
\delta \Delta_n^{{\rm HREC},a}  &=
\frac{(\Za)^3}{\pi} \, \frac{m}{M}\, E_F\,\ln (\Za)\,
  \nonumber \\ & \times
   \left(-\frac43\right)\,\left[ 1-\frac1n+\gamma+\Psi(n)-\ln(n)
     \right]\,.
\end{align}
This expression generalizes the result for the difference
$\Delta_2$ reported in Ref.~\cite{KaIv2002plb}. The second
contribution (absent in Ref.~\cite{KaIv2002plb}) is obtained as a
second-order perturbation induced by the operator $H_S$ and by the
operator responsible for the {\it nonlogarithmic} recoil
correction to the Lamb shift to order $(\Za)^5\,m^2/M$. The
logarithm of $\Za$ then arises from the second term of the $\Za$
expansion of the electron propagator after an integration over the
logarithmic region \cite{karshenboim:93:jetp}. The result reads
\begin{align} \label{HRECb}
\delta \Delta_n^{{\rm HREC},b}  &=
\frac{(\Za)^3}{\pi} \, \frac{m}{M}\, E_F\, \ln (\Za)\,
  \nonumber \\ & \times
   \frac{28}{3} \left[ - \frac12 + \frac{1}{2 n}+
                     \gamma + \Psi(n) - \ln(n)
     \right]\,.
\end{align}
We note that this contribution, unlike Eq.~(\ref{HRECa}), is
finite for single $nS$ states. For $1S$ state, the constant in
Eq.~(\ref{HRECb}) turns into $(124/9+28/3\,\ln 2)$, which
coincides with a part of the complete $1S$ result obtained by
Kinoshita \cite{Ki1998} ($2\,C_S$ in his notation). Our result
for the logarithmic part of the fourth-order recoil correction is
the sum of Eqs.~(\ref{HRECa}) and~(\ref{HRECb}),
\begin{align} \label{logrec}
\delta \Delta_n^{\rm HREC} = & \;
\frac{(Z\alpha)^3}{\pi} \,
\frac{m}{M}\, E_F \,\ln (\Za)\,
\nonumber\\
& \times  8 \,\left[- \frac34+ \frac{3}{4n}  + \gamma + \Psi (n) -
\ln(n) \right]\,.
\end{align}
We do not have a proof that this result is complete.

Some {\em incomplete} results for the fourth-order one-loop
self-energy and vacuum-polarization corrections were obtained in
Ref.~\cite{Ka2001hyp}. With misprints being corrected in
\cite{KaIv2002}, these corrections read, respectively,
\begin{subequations}
\begin{align} \label{se:ho}
&  \delta \Delta_n^{\rm HSE} = \alpha (Z\alpha)^3 E_F
 \left[ -\frac{621}{320}\frac{n^2-1}{n^2}
   + \left( \frac{191}{16} - 5\ln 2 \right) \,
     \right. \nonumber\\ & \quad \left. \times
   \left(\frac{11}{20} -\frac{1}{n} +\frac{9}{20\,n^2} +
   \gamma   + \Psi (n) - \ln(n) \right)
   \right]\,,
\\[2em]
\label{vp:ho}
&  \delta \Delta_n^{\rm HVP}  = \alpha \, (Z\alpha)^3 \, E_F\,
\left( -\frac{13}{24}\right) \,
     \nonumber\\ & \quad \times
 \left[
 -\frac{55}{26} -\frac1n + \frac{81}{26\,n^2}+
\gamma  + \Psi (n)  - \ln(n) \right]\,.
\end{align}
\end{subequations}
It should be noted that the one-loop self-energy correction yields
the largest contribution among all fourth-order corrections
mentioned so far and the incompleteness of the result
(\ref{se:ho}) provides the dominant theoretical uncertainty 
for $\Delta_n$. For the particular case $n=2$, this
correction was evaluated numerically to all orders in $\Za$ in
Refs.~\cite{YeSh2001hyp,YeArShPl2005}. The deviation of the
contribution (\ref{se:ho}) from the all-order result was found 
to be on
the level of 20\%. The evaluation of the complete result for the
fourth-order vacuum-polarization correction is a much simpler task
than for the self-energy. It can be solved either analytically, as
was done for $n=2$ in Ref.~\cite{karshenboim:00:jetp,KaIv2002}, or
(which is much easier) numerically, as was done for $n=2$ in
Ref.~\cite{YeArShPl2005}. However, in view of the absence of complete
results for the self-energy correction, we do not pursue the
matter any further in the current investigation.

The nuclear-structure correction was found in
Refs.~\cite{Ka1997nuc,Ka2001hyp} to be
\begin{align}   \label{nucl}
&  \delta \Delta_n^{\rm Nucl}  = -(\Za)^2\,
 \Delta E_{\rm 1S}^{\rm Nucl}\,
 \left[ - \frac{5}{4} - \frac{1}{n} + \frac{9}{4\,n^2}  + \gamma
     \right. \nonumber\\ & \quad
  + \Psi (n) -
  \ln (n)\Biggr]
+ \frac43\,(\Za)^2\, \Biggl[ \gamma+\Psi(n)-\ln(n)
     \ \nonumber\\ & \quad \left.
 + \frac{n-1}{n}
 -\left(\frac{R_M}{R_E}\right)^2\, \frac{n^2-1}{4\,n^2}\right]
    \left( m\,R_E \right)^2\,E_F\,,
\end{align}
where $R_E$ and $R_M$ are the electric and the magnetic charge
radii, respectively, and $\Delta E_{\rm 1S}^{\rm Nucl}$ is the
nuclear correction for the ground-state hfs.

%
%
\section{Theoretical results for $\Delta_n$} \label{subsec43}

In this section, we collect all theoretical contributions available
to the normalized difference of $nS$ states $\Delta_n$
[Eq.~(\ref{Delta})]. Numerical results for individual
contributions and the total theoretical values of $\Delta_n$ in
hydrogen  are listed in Table~\ref{results} for principal quantum
numbers $n = 1,\dots,8$. The second- and third-order corrections
summarized by Eq.~(\ref{the1}) are given in the first five rows of
this Table. Forth-order QED corrections discussed in
Sec.~\ref{subsec42} are tabulated in the next seven rows, and the
nuclear-structure correction completes the analysis. Parameters of
the proton used for calculating numerical data in
Table~\ref{results} agree with those from Table~8 of
Ref.~\cite{Ka2005}. The nuclear-structure correction for the
ground-state hfs that enters Eq.~(\ref{nucl}) was taken from
Ref.~\cite{KaIv2002}, where it was obtained by subtracting all
known QED corrections from the experimental result for the
ground-state hfs (\ref{E1S}). Its numerical value is $-46\,{\rm
kHz}$.

We already mentioned above that in the particular case $n=2$,
there are complete all-order results available
for the $\delta \Delta_n^{\rm HSE}$ and $\delta \Delta_n^{\rm HVP}$ 
corrections. We thus employ the numerical
values for the self-energy and vacuum-polarization remainder
functions for the difference $\Delta_2$ as given in
Ref.~\cite{YeArShPl2005}, as well as the uncertainty estimates
given in the cited reference. The corresponding entries in the
table are marked with the asterisk. For $n>2$, we use the formulas
(\ref{se:ho}) and~(\ref{vp:ho}) and ascribe the 50\% uncertainty
to them. The error estimates for the other forth-order corrections
are as follows: for the two-loop and the radiative recoil
corrections, we assume the uncertainty to be a half the numerical
value of the logarithmic terms, while for the recoil correction we
use 100\% of the correction given by Eq.~(\ref{logrec}).

The two last rows of Table~\ref{results} are reserved for the
total theoretical predictions for the normalized difference
$\Delta_n$ and for the complete values of the hfs frequency of
excited hydrogenic $nS$ states. The latter are obtained by
combining the highly accurate 
experimental value of the ground-state hfs interval
(\ref{E1S}) and the theoretical prediction for $\Delta_n$ given in
the previous row of the table.

For the case $n=2$, our evaluation differs from the previous
investigation of the difference $\Delta_2$ presented in
Ref.~\cite{KaIv2002} in two ways: (i) we employ the latest
numerical results for the self-energy remainder from
Ref.~\cite{YeArShPl2005} and the error estimate from this
reference and (ii) we also have found
an additional (numerically small) higher-order
logarithmic recoil contribution (\ref{HRECb}).
Despite the small change of the theoretical
prediction, our final result for the hfs frequency of the $2S$
state still deviates by $1.4\,\sigma$ from the experimental result
$E_{2S} = 177\,566\,860(16)\,$~Hz \cite{KoFiKaHa2004}. We mention
also a similar ($1.8\,\sigma$) deviation of the theoretical value
of $\Delta_2$ for the $^3$He ion from the experimental result
(\ref{1}) observed in Ref.~\cite{YeArShPl2005}.

%
%
\section{CONCLUSION}
\label{sec5}

The normalized difference of the hfs intervals $\Delta_{2} = 8 \,
\Delta E_{2S} - \Delta E_{1S}$ has been a subject for both
theoretical and experimental investigations since a long time. In
this paper, we have presented calculations that generalize the
previous studies of $\Delta_{n} = n^3 \,
\Delta E_{nS} - \Delta E_{1S}$ to general $n$. Our results are complete through
third order in the parameters $\alpha$, $\Za$, and $m/M$; an
estimation of the fourth-order corrections is also supplied.

The dominant source of the present theoretical uncertainty for the
difference $\Delta_{n}$ comes from the higher-order one-loop
self-energy correction. Further improvement of the theory can be
achieved by a numerical all-order (in $\Za$) evaluation of this
correction. Such a calculation has been carried out for the difference
$\Delta_{2}$ in Refs.~\cite{YeSh2001hyp,YeArShPl2005} based on a
method developed by a number of authors
\cite{blundell:97:prl,yerokhin:97:eprint,sunnergren:98:pra} and
seems feasible for higher values of $n$ as well. It should be
noted that the results for hydrogen reported in
Refs.~\cite{YeSh2001hyp,YeArShPl2005} involved an extrapolation of
numerical data obtained for $Z\ge 5$ towards $Z=1$. It would
clearly be preferable to perform a direct numerical calculation of
the higher-order self-energy correction for $Z=1$, as it was done
for the Lamb shift in Refs.~\cite{JeMoSo1999,JeMo2004pra}. This
project is underway.

%
%
\section*{Acknowledgments}

The authors acknowledge helpful discussions with P.~J.~Mohr.
U.D.J.~acknowledges support from Deutsche Forschungsgemeinschaft
(DFG, Heisenberg program) under contract JE285/3-1, and
V.A.Y.~gratefully acknowledges support from RFBR under contract
04-02-17574. This project has also been supported by the DFG
collaborative research grant 436 RUS 113/853/0-1.

\begin{widetext}

\begin{table}[htb]
\caption{\label{results} Individual contributions to the
normalized difference $\Delta_n$ of hfs frequencies, and absolute
values of the hyperfine splitting frequencies of excited $S$
states in hydrogen. For the entries marked with an asterisk
($^*$), we employ the numerical results for the self-energy and
vacuum-polarization remainder functions as reported in
Ref.~\cite{YeArShPl2005} instead of the analytic expressions given
in Eqs.~(\ref{se:ho}) and~(\ref{vp:ho}) used in other cases. The
absolute values for the hfs frequencies of excited states are
obtained with the help of $1S$ experimental result in
Eq.~(\ref{E1S}) as a reference. Units are Hz.}
\begin{scriptsize}
\begin{tabular}{cc@{\hspace{0.4cm}}c@{\hspace{0.4cm}}c@{\hspace{0.4cm}}%
c@{\hspace{0.4cm}}c@{\hspace{0.4cm}}c@{\hspace{0.4cm}}c}
\hline
\hline
\rule[-3mm]{0mm}{8mm}
Effect &
\multicolumn{1}{c}{$2S$} &
\multicolumn{1}{c}{$3S$} &
\multicolumn{1}{c}{$4S$} &
\multicolumn{1}{c}{$5S$} &
\multicolumn{1}{c}{$6S$} &
\multicolumn{1}{c}{$7S$} &
\multicolumn{1}{c}{$8S$} \\
\hline
\rule[-2mm]{0mm}{6mm} $(\Za)^2$ & 47~222.0 & 47~571.8 & 44~860.9 & 42~310.9 & 40~226.1 & 38~548.6 & 37~187.3\\
\rule[-2mm]{0mm}{6mm} $\alpha\,(\Za)^2$ (SE) & 1~936.0 & 2~718.6 & 3~134.2 & 3~390.9 & 3~564.9 & 3~690.4 & 3~785.3\\
\rule[-2mm]{0mm}{6mm} $\alpha\,(\Za)^2$ (VP) & $-$58.0 & $-$79.2 & $-$90.1 & $-$96.8 & $-$101.3 & $-$104.5 & $-$106.9\\
\rule[-2mm]{0mm}{6mm} $(\Za)^2\,(m/M)$ & $-$162.9 & $-$210.3 & $-$232.6 & $-$245.6 & $-$254.0 & $-$260.0 & $-$264.4\\
\hline
\rule[-2mm]{0mm}{6mm} Sum of 3$^{\rm rd}$ order & 48~937.1 & 50~000.9 & 47~672.4 & 45~359.4 & 43~435.7 & 41~874.5 & 40~601.3\\
\hline
\hline
\rule[-2mm]{0mm}{6mm} $(\Za)^4$ & 5.6 & 5.6 & 5.2 & 4.9 & 4.6 & 4.4 & 4.2\\
\rule[-2mm]{0mm}{6mm} $\alpha^2\,(\Za)^2$ & 3.3(1.7) & 4.5(2.3) & 5.1(2.6) & 5.5(2.8) & 5.8(2.9) & 6.0(3.0) & 6.1(3.1)\\
\rule[-2mm]{0mm}{6mm} $\alpha\,(\Za)^2\,(m/M)$ & $-$3.1(1.6) & $-$4.2(2.1) & $-$4.8(2.4) & $-$5.2(2.6) & $-$5.4(2.7) & $-$5.6(2.8) & $-$5.7(2.9)\\
\rule[-2mm]{0mm}{6mm} $\alpha\,(\Za)^3$ (SE) & 9.7(5)$^*$ & 15.8(7.9) & 19.1(9.6) & 21.2(10.6) & 22.7(11.3) & 23.7(11.9) & 24.5(12.3)\\
\rule[-2mm]{0mm}{6mm} $\alpha\,(\Za)^3$ (VP) & 3.0$^*$ & 3.7(1.9) & 3.8(1.9) & 3.7(1.9) & 3.7(1.9) & 3.7(1.8) & 3.7(1.8)\\
\rule[-2mm]{0mm}{6mm} $(\Za)^3\,(m/M)$ & 0.3(3) & 0.4(4) & 0.4(4) & 0.5(5) & 0.5(5) & 0.5(5) & 0.5(5)\\
\hline
\rule[-2mm]{0mm}{6mm} Sum of 4$^{\rm th}$ order & 18.7(2.3) & 25.8(8.7) & 28.8(10.4) & 30.6(11.4) & 31.8(12.2) & 32.7(12.7) & 33.3(13.1)\\
\hline
\hline
\rule[-2mm]{0mm}{6mm} Nucl & $-$1.8 & $-$1.8 & $-$1.7 & $-$1.6 & $-$1.5 & $-$1.5 & $-$1.4\\
\hline
\hline
\rule[-2mm]{0mm}{6mm} Total $\Delta_n$ & 48~954.0(2.3) & 50~024.9(8.7) & 47~699.5(10.4) & 45~388.4(11.4) & 43~466.0(12.2) & 41~905.7(12.7) & 40~633.2(13.1)\\
\hline
\rule[-2mm]{0mm}{6mm} HFS freq. & 177~556~838.2(3)  & 52~609~473.2(3)  & 22~194~585.2(2)  & 11~363~609.1(1)  & 6~576~153.79(6)  & 4~141~246.81(4)  & 2~774~309.35(3)  \\
\hline
\hline
\end{tabular}
\end{scriptsize}
\end{table}

\end{widetext}

%
%


\begin{thebibliography}{10}

\bibitem{Ka2005}
S.~G. Karshenboim, Phys. Rep. {\bf 422},  1  (2005).

\bibitem{EsDoBaHo1971}
L. Essen, R.~W. Donaldson, M.~J. Bangham, and E.~G. Hope, Nature (London) {\bf
  229},  110  (1971);
L. Essen, R.~W. Donaldson, E.~G. Hope, and M.~J. Bangham, Metrologia {\bf 9},
  128  (1973).

\bibitem{St1963}
M.~M. Sternheim, Phys. Rev. {\bf 130},  211  (1963).

\bibitem{KoFiKaHa2004}
N. Kolachevsky, M. Fischer, S.~G. Karshenboim, and T.~W. H\"{a}nsch, Phys. Rev.
  Lett. {\bf 92},  033003  (2004).

\bibitem{KoFeKaHa2004}
N. Kolachevsky, P. Fendel, S.~G. Karshenboim, and T.~W. H\"{a}nsch, Phys. Rev.
  A {\bf 70},  062503  (2004).

\bibitem{ScFoDe1969}
H.~A. Schluessler, E.~N. Fortson, and H.~G. Dehmelt, Phys. Rev. {\bf 187},  5
  (1969), [Erratum Phys. Rev. A {\bf 2}, 1612 (E) (1970)].

\bibitem{PrWa1977}
M.~H. Prior and E.~C. Wang, Phys. Rev. A {\bf 16},  6  (1977).

\bibitem{Zw1961}
D. Zwanziger, Phys. Rev. {\bf 121},  1128  (1961).

\bibitem{UdPriv2005}
{Th.~Udem}, private communication (2005).

\bibitem{ArPriv2004}
O. Arnoult, private communication (2004).

\bibitem{Pa1995}
K. Pachucki, Phys. Rev. A {\bf 53},  2092  (1995).

\bibitem{NiKi1997}
M. Nio and T. Kinoshita, Phys. Rev. D {\bf 55},  7267  (1997).

\bibitem{Je2003jpa}
U.~D. Jentschura, J. Phys. A {\bf 36},  L229  (2003).

\bibitem{ItZu1980}
C. Itzykson and J.~B. Zuber, {\em Quantum Field Theory} (McGraw-Hill, New York,
  NY, 1980).

\bibitem{B5}
For the first time, the coefficient $a_{20}^{\rm SE}(2,1)$ was
evaluated by Zwanziger \cite{Zw1961} to be $a_{20}^{\rm SE}(2,1) =
-5.37(6)$. A more accurate value for this coefficient was later
obtained by P. J. Mohr by recalculating the integrals listed in
Eq.~(B.5) of Ref.~\cite{Zw1961} (private communication). According to
P. J. Mohr, the expression $(1-s)^2$ in Eq.~(B.5) of the cited
reference should be replaced by $(1-s)$. After the elimination of
this typographical error, the formulas of Ref.~\cite{Zw1961} may
be used for an accurate evaluation of the difference $a_{20}^{\rm
SE}(2,1)$. The private communication by P. J. Mohr is also 
quoted as reference number~[18] of Ref. \cite{PrWa1977}, and the
value of $-5.5515$ given in Eq.~(37) of Ref.~\cite{PrWa1977} is
the sum of $a_{20}^{\rm SE}(2,1) + a_{20}^{\rm VP}(2,1)$,
which implies a value of $-5.2212$ for $a_{20}^{\rm SE}(2,1)$.
According to S. G. Karshenboim (private communication),
the value of $-5.221\,233(3)$ for $a_{20}^{\rm SE}(2,1)$ has been obtained
independently by J. R. Sapirstein and S. G. Karshenboim in an
unpublished investigation, as cited in Ref.~\cite{Ka2001hyp}.

\bibitem{Ka2001hyp}
S.~G. Karshenboim,  in {\em The Hydrogen Atom -- Lecture Notes in Physics
  Vol.~570}, edited by S.~G. Karshenboim and F.~S. Pavone (Springer, Berlin,
  2001), pp.\ 335--343.

\bibitem{JeCzPa2005}
U.~D. Jentschura, A. Czarnecki, and K. Pachucki, Phys. Rev. A {\bf 72},  062102
   (2005).

\bibitem{PrBrMa2002}
A.~P. Prudnikov, Yu.~A. Brychkov, and O.~I. Marychev, {\em Integrals and Sums},
  2 ed. (Fizmatlit, Moscow, 2002), in Russian.

\bibitem{Br1930}
G. Breit, Phys. Rev. {\bf 35},  1477  (1930).

\bibitem{Ka1997nuc}
S.~G. Karshenboim, Phys. Lett. A {\bf 225},  97  (1997).

\bibitem{YeSh2001hyp}
V.~A. Yerokhin and V.~M. Shabaev, Phys. Rev. A {\bf 64},  012506  (2001).

\bibitem{KaIv2002}
S.~G. Karshenboim and V.~G. Ivanov, Eur. Phys. J. D {\bf 19},  13  (2002).

\bibitem{KaIv2002plb}
S.~G. Karshenboim and V.~G. Ivanov, Phys. Lett. B {\bf 524},  259  (2002).

\bibitem{KaIv2005}
S.~G. Karshenboim and V.~G. Ivanov, Can. J. Phys. {\bf 83},  1063  (2005).

\bibitem{YeArShPl2005}
V.~A. Yerokhin, A.~N. Artemyev, V.~M. Shabaev, and G. Plunien,
Phys. Rev. A  {\bf 72},  052510  (2005).

\bibitem{karshenboim:93:jetp}
S.~G. Karshenboim, Zh. \'{E}ksp. Teor. Fiz. {\bf 103}, 1105 (1993)
[JETP {\bf 76}, 541 (1993)].

\bibitem{Ki1998}
T. Kinoshita, e-print hep-ph/9808351 (1998).

\bibitem{karshenboim:00:jetp}
S.~G. Karshenboim, V.~G. Ivanov, and V.~M. Shabaev,
\newblock Zh. \'Eksp. Teor. Fiz. {\bf 117}, 67  (2000)
\newblock [JETP {\bf 90}, 59 (2000)].

\bibitem{blundell:97:prl}
S.~A. Blundell, K.~T. Cheng, and J.~Sapirstein,
\newblock Phys. Rev. Lett. {\bf 78}, 4914 (1997).

\bibitem{yerokhin:97:eprint}
V.~A. Yerokhin, V.~M. Shabaev, and A.~N. Artemyev,
\newblock e-print physics/9705029 (1997).

\bibitem{sunnergren:98:pra}
P.~Sunnergren, H.~Persson, S.~Salomonson, S.~M. Schneider,
I.~Lindgren, and
  G.~Soff,
\newblock Phys. Rev. A {\bf 58}, 1055 (1998).

\bibitem{JeMoSo1999}
U.~D. Jentschura, P.~J. Mohr, and G. Soff, Phys. Rev. Lett. {\bf 82},  53
  (1999).

\bibitem{JeMo2004pra}
U.~D. Jentschura and P.~J. Mohr, Phys. Rev. A {\bf 69},  064103  (2004).

\end{thebibliography}
\end{document}